\PassOptionsToPackage{unicode}{hyperref}
\PassOptionsToPackage{hyphens}{url}
\PassOptionsToPackage{dvipsnames,svgnames,x11names}{xcolor}
\documentclass[
]{article}

\usepackage{amsmath,amssymb}
\usepackage{iftex}
\ifPDFTeX
  \usepackage[T1]{fontenc}
  \usepackage[utf8]{inputenc}
  \usepackage{textcomp} 
\else 
  \usepackage{unicode-math}
  \defaultfontfeatures{Scale=MatchLowercase}
  \defaultfontfeatures[\rmfamily]{Ligatures=TeX,Scale=1}
\fi
\usepackage{lmodern}
\ifPDFTeX\else  
\fi
\IfFileExists{upquote.sty}{\usepackage{upquote}}{}
\IfFileExists{microtype.sty}{
  \usepackage[]{microtype}
  \UseMicrotypeSet[protrusion]{basicmath} 
}{}
\makeatletter
\@ifundefined{KOMAClassName}{
  \IfFileExists{parskip.sty}{%
    \usepackage{parskip}
  }{
    \setlength{\parindent}{0pt}
    \setlength{\parskip}{6pt plus 2pt minus 1pt}}
}{
  \KOMAoptions{parskip=half}}
\makeatother
\usepackage{xcolor}
\ifx\paragraph\undefined\else
  \let\oldparagraph\paragraph
  \renewcommand{\paragraph}[1]{\oldparagraph{#1}\mbox{}}
\fi
\ifx\subparagraph\undefined\else
  \let\oldsubparagraph\subparagraph
  \renewcommand{\subparagraph}[1]{\oldsubparagraph{#1}\mbox{}}
\fi

\usepackage{color}
\usepackage{fancyvrb}

\DefineVerbatimEnvironment{Highlighting}{Verbatim}{commandchars=\\\{\}}
\usepackage{framed}
\definecolor{shadecolor}{RGB}{241,243,245}
\newenvironment{Shaded}{\begin{snugshade}}{\end{snugshade}}

\newcommand{\BuiltInTok}[1]{\textcolor[rgb]{0.00,0.23,0.31}{#1}}

\newcommand{\CommentTok}[1]{\textcolor[rgb]{0.37,0.37,0.37}{#1}}

\newcommand{\ControlFlowTok}[1]{\textcolor[rgb]{0.00,0.23,0.31}{#1}}

\newcommand{\DecValTok}[1]{\textcolor[rgb]{0.68,0.00,0.00}{#1}}

\newcommand{\ErrorTok}[1]{\textcolor[rgb]{0.68,0.00,0.00}{#1}}

\newcommand{\FloatTok}[1]{\textcolor[rgb]{0.68,0.00,0.00}{#1}}

\newcommand{\ImportTok}[1]{\textcolor[rgb]{0.00,0.46,0.62}{#1}}

\newcommand{\KeywordTok}[1]{\textcolor[rgb]{0.00,0.23,0.31}{#1}}
\newcommand{\NormalTok}[1]{\textcolor[rgb]{0.00,0.23,0.31}{#1}}
\newcommand{\OperatorTok}[1]{\textcolor[rgb]{0.37,0.37,0.37}{#1}}

\newcommand{\SpecialStringTok}[1]{\textcolor[rgb]{0.13,0.47,0.30}{#1}}
\newcommand{\StringTok}[1]{\textcolor[rgb]{0.13,0.47,0.30}{#1}}

\usepackage{longtable,booktabs,array}
\usepackage{calc} 
\usepackage{etoolbox}
\makeatletter
\patchcmd\longtable{\par}{\if@noskipsec\mbox{}\fi\par}{}{}
\makeatother
\IfFileExists{footnotehyper.sty}{\usepackage{footnotehyper}}{\usepackage{footnote}}
\makesavenoteenv{longtable}
\usepackage{graphicx}
\makeatletter
\def\maxwidth{\ifdim\Gin@nat@width>\linewidth\linewidth\else\Gin@nat@width\fi}
\def\maxheight{\ifdim\Gin@nat@height>\textheight\textheight\else\Gin@nat@height\fi}
\makeatother
\setkeys{Gin}{width=\maxwidth,height=\maxheight,keepaspectratio}
\makeatletter
\def\fps@figure{htbp}
\makeatother
\newlength{\cslhangindent}
\newlength{\csllabelwidth}
\newlength{\cslentryspacingunit} 
\newenvironment{CSLReferences}[2] 
 {
  \setlength{\parindent}{0pt}
  \ifodd #1
  \let\oldpar\par
  \def\par{\hangindent=\cslhangindent\oldpar}
  \fi
  \setlength{\parskip}{#2\cslentryspacingunit}
 }%
 {}
\usepackage{calc}

\newcommand{\CSLLeftMargin}[1]{\parbox[t]{\csllabelwidth}{#1}}
\newcommand{\CSLRightInline}[1]{\parbox[t]{\linewidth - \csllabelwidth}{#1}\break}

\usepackage{arxiv}
\usepackage{orcidlink}
\usepackage{amsmath}
\usepackage[T1]{fontenc}
\makeatletter
\@ifpackageloaded{caption}{}{\usepackage{caption}}
\AtBeginDocument{%
\ifdefined\contentsname
  \renewcommand*\contentsname{Table of contents}
\else
  \newcommand\contentsname{Table of contents}
\fi
\ifdefined\listfigurename
  \renewcommand*\listfigurename{List of Figures}
\else
  \newcommand\listfigurename{List of Figures}
\fi
\ifdefined\listtablename
  \renewcommand*\listtablename{List of Tables}
\else
  \newcommand\listtablename{List of Tables}
\fi
\ifdefined\figurename
  \renewcommand*\figurename{Figure}
\else
  \newcommand\figurename{Figure}
\fi
\ifdefined\tablename
  \renewcommand*\tablename{Table}
\else
  \newcommand\tablename{Table}
\fi
}
\@ifpackageloaded{float}{}{\usepackage{float}}
\floatstyle{ruled}
\@ifundefined{c@chapter}{\newfloat{codelisting}{h}{lop}}{\newfloat{codelisting}{h}{lop}[chapter]}
\floatname{codelisting}{Listing}

\makeatother
\makeatletter
\@ifpackageloaded{caption}{}{\usepackage{caption}}
\@ifpackageloaded{subcaption}{}{\usepackage{subcaption}}
\makeatother
\makeatletter
\@ifpackageloaded{tcolorbox}{}{\usepackage[skins,breakable]{tcolorbox}}
\makeatother
\makeatletter
\@ifundefined{shadecolor}{\definecolor{shadecolor}{rgb}{.97, .97, .97}}
\makeatother
\ifLuaTeX
  \usepackage{selnolig}  
\fi
\IfFileExists{bookmark.sty}{\usepackage{bookmark}}{\usepackage{hyperref}}
\IfFileExists{xurl.sty}{\usepackage{xurl}}{} 
\hypersetup{
  pdftitle={Slicenet: a Simple and Scalable Flow-Level Simulator for Network Slice Provisioning and Management},
  pdfauthor={Viswanath KumarSkandPriya; Abdulhalim Dandoush; Gladys Diaz},
  pdfkeywords={Network Slicing, 5G, Software Defined Networks, Network
Functions Virtualization, Simulators},
  colorlinks=true,
  linkcolor={blue},
  filecolor={Maroon},
  citecolor={Blue},
  urlcolor={Blue},
  pdfcreator={LaTeX via pandoc}}

\newcommand{\runninghead}{A Preprint }
\renewcommand{\runninghead}{for arXiv }
\title{Slicenet: a Simple and Scalable Flow-Level Simulator for Network
Slice Provisioning and Management}
\def\asep{\\\\\\ } 
\author{\textbf{Viswanath KumarSkandPriya}\\Esme research lab, Campus
Paris Sud, France\\\\\asep\textbf{Abdulhalim Dandoush}\\University of
Doha for Science and Technology, Doha, Qatar\\\\Esme research lab,
Campus Paris Sud, France\\\\\asep\textbf{Gladys Diaz}\\Universite
Sorbonne Paris Nord - L2TI, 93430 Villateneuse, France\\\\}
\date{}
\begin{document}
\maketitle
\begin{abstract}
Network slicing plays a crucial role in the progression of 5G and
beyond, facilitating dedicated logical networks to meet diverse and
specific service requirements. The principle of End-to-End (E2E) slice
includes not only a service chain of physical or virtual functions for
the radio and core of 5G/6G networks but also the full path to the
application servers that might be running at some edge computing or at
central cloud.

Nonetheless, the development and optimization of E2E network slice
management systems necessitate a reliable simulation tool for evaluating
different aspects at large-scale network topologies such as resource
allocation and function placement models. This paper introduces
Slicenet, a mininet-like simulator crafted for E2E network slicing
experimentation at the flow level.

Slicenet aims at facilitating the investigation of a wide range of slice
optimization techniques, delivering measurable, reproducible results
without the need for physical resources or complex integration tools. It
provides a well-defined process for conducting experiments, which
includes the creation and implementation of policies for various
components such as edge and central cloud resources, network functions
of multiple slices of different characteristics. Furthermore, Slicenet
effortlessly produces meaningful visualizations from simulation results,
aiding in comprehensive understanding.

Utilizing Slicenet, service providers can derive invaluable insights
into resource optimization, capacity planning, Quality of Service (QoS)
assessment, cost optimization, performance comparison, risk mitigation,
and Service Level Agreement (SLA) compliance, thereby fortifying network
resource management and slice orchestration.
\end{abstract}
{\bfseries \emph Keywords}
\def\sep{\textbullet\ }
Network Slicing \sep 5G \sep Software Defined Networks \sep Network
Functions Virtualization \sep 
Simulators

\ifdefined\Shaded\renewenvironment{Shaded}{\begin{tcolorbox}[sharp corners, enhanced, interior hidden, borderline west={3pt}{0pt}{shadecolor}, boxrule=0pt, frame hidden, breakable]}{\end{tcolorbox}}\fi

\hypertarget{sec-intro}{%
\section{Introduction}\label{sec-intro}}

The advent of cutting-edge network technologies such as Network Function
Virtualization (NFV) and Software-Defined Networking (SDN) has
substantially reshaped the landscape of telecommunications and cloud
computing {[}1--3{]}. These technologies offer the capability for
dynamic resource allocation and sophisticated management of network
services, pivotal components for actualizing network slicing - a
cornerstone feature for 5G and beyond {[}1, 4{]}. Despite its promising
potential, efficient orchestration and management of network slices for
optimized resource admission and allocation pose substantial challenges
{[}4--6{]}.

One of the key concerns in the 5G and beyond era is the promise of
end-to-end network management. It's not just about ensuring seamless
network performance, but also about delivering on the promise of
fairness between users of similar profiles {[}1{]}. This capability is
imperative for accommodating new generation applications and use-cases
like tele-surgery, online education using mixed reality and metaverse,
and bigdata applications with dedicated slices for predefined user
requirements{[}7{]}. For these applications, not only are high-speed and
low-latency connectivity crucial, but so too are guaranteed Quality of
Service (QoS), high availability, and reliability. The network slices
dedicated to these applications should be handled with advanced
management and orchestration tools to ensure fair and optimized resource
allocation.

This is where Slicenet steps in. It's a specialized simulator tailored
for large-scale network topologies, incorporating User Equipment (UEs),
Application nodes, Access networks, Transport networks, Core network,
and Data network. Slicenet offers researchers a comprehensive
environment to experiment with various network slicing optimizations,
including Slice Orchestration, Slice Admission, Slice Service-Level
Agreement (SLA) Adherence, Violation Avoidance, Slice Composition, and
Dynamic Slicing Negotiation techniques {[}3{]}. Slicenet consistently
delivers measurable and reproducible results, offering enhanced insights
into resource optimization, scheduling, prioritization, and capacity
models. One of its unique capabilities is its ability to abstract
traffic patterns from topology, allowing the same topology to be
evaluated under different traffic scenarios {[}8{]}. By enabling
researchers and service providers to thoroughly understand and optimize
network slicing management, Slicenet emerges as a critical tool in the
5G and beyond landscape. It stands as a vanguard in the pursuit of
overcoming the challenges and realizing the full potential of network
slicing for an array of next-generation applications and use cases.

The rest of this paper is organized as follows. Section~\ref{sec-2}
overviews the related works and highlights the distinct contribution of
this work. The Architecture of Slicenet is provided in
Section~\ref{sec-3}. Section~\ref{sec-4} will discuss how to run a
complete E2E simulation scenario. In Section~\ref{sec-5}, we show how to
easily visualize the results in understandable way. Last,
Section~\ref{sec-6} concludes our works and provides some future
directions.

\hypertarget{sec-2}{%
\section{Related Works}\label{sec-2}}

Current network simulation tools such as Mininet {[}9{]} and its
derivatives, including Mininet-Wifi {[}10{]} and Containernet {[}11{]},
focus on emulating real-world devices, switches, hosts, and their
interactions in specific topologies, with a focus on Software-Defined
Networking (SDN) and Network Functions Virtualization (NFV).

On the other hand, CloudSim and its offshoots, like EdgeCloudSim
{[}12{]} and CloudSimHypervisor {[}13{]}, address cloud-specific
challenges around data center optimization and resource provisioning on
compute and network infrastructure. However, these tools either overlook
the impacts of network slicing and related Quality of Service (QoS) or
reduce the complexity of the network slicing problem, thereby failing to
address critical issues in the field.

Unlike its predecessors, Slicenet efficiently simulates network
topologies and application traffic at the flow level {[}8{]}, reducing
the necessity for extensive emulation of physical or virtual topologies.
This characteristic opens the door to a more consistent and simpler
experimentation process for both operators and researchers and enables
the evaluation of large scale deployment of a particular algorithm
(e.g., AI-based resource allocation or NFV placement model). Building
upon the successful simulation concepts from previous tools, Slicenet
offers an integrated environment for the experimentation and
documentation of findings. This tool is Python-based, lending itself
seamlessly to the use of popular Machine Learning frameworks such as
TensorFlow and PyTorch, to use Machine Learning, Deep Learning, and
Neural Network models as part of the experiments. Furthermore, the
simulations can be run in a Jupyter notebook setup without requiring
complex computing or networking requirements thereby promoting swift
sharing and visualization of results, thereby fostering wider community
collaboration.

\hypertarget{sec-3}{%
\section{Architecture of Slicenet}\label{sec-3}}

\begin{figure}

{\centering \includegraphics{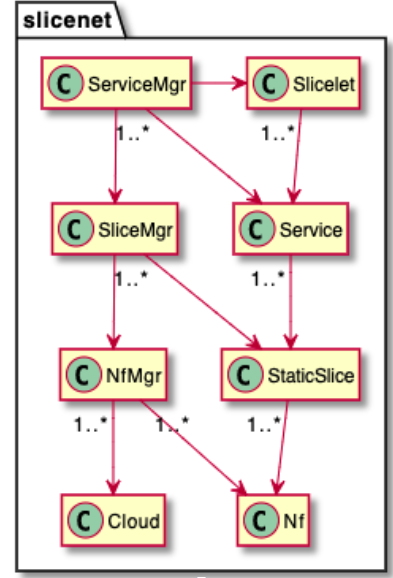}

}

\caption{\label{fig-slicenet-arch}slicenet available as python package}

\end{figure}

The first iteration of slicenet is introduced as a python package
Figure~\ref{fig-slicenet-arch} containing basic foundational classes for
expressing a networking slicing infrastructure topology and to model
user traffic patterns.

\hypertarget{base-entities}{%
\subsection*{BASE ENTITIES}\label{base-entities}}
\addcontentsline{toc}{subsection}{BASE ENTITIES}

\texttt{Cloud} class abstracts the overall network slice infrastructure
as the unit of compute, memory and storage units. In future revisions of
this class, other nuances like location, networking characteristics,
HPA/VPA characteristics, application constraints and dependencies can be
added to further enrich the representation of infrastructure as needed.

\texttt{Nf} class abstracts the network function resource requirements
as the unit of percentage of compute, memory and storage. This class
also provides the capability to report the what slices are being powered
by this network function, how much of this network function's resources
are being utilized by what slice (as a percentage factor) and what is
the overall utilization ratio of this network function with respect to
overlying slices.

\texttt{StaticSlice} class abstracts the StaticSlice resource
requirements as the unit of percentage share of the underlying
\texttt{Nf} objects. An E2E Slice is being composed of underlying
network functions and service chains. There is a \emph{M:N relationship}
between a slice and network function object. A Slice can be composed of
multiple Network Functions and a Network Function can power multiple
slices. Inorder to depict this \emph{M:N relationship}, this
\texttt{StaticSlice} class contains the percentage allocation of
underlying Nf objects in its definition. In future versions of slicenet,
DynamicSlices will be introduced. These slices are created by the slice
management system on demand based on the traffic pattern, load and
estimations. Dynamic slices can be either elastic or non-elastic in
nature.

\texttt{Service} class abstracts the Communication Service requirements
as the unit of percentage share of the underlying StaticSlice objects
along with a priority clause. \texttt{StaticSlice} as the name suggests,
is static in nature and powers the overlying services. Hence by virtue,
StaticSlice(s) are being Shared by the Service(s)

\hypertarget{slicelet}{%
\subsubsection*{SLICELET}\label{slicelet}}
\addcontentsline{toc}{subsubsection}{SLICELET}

Slicenet follows a specific model for user traffic simulation known as
the \texttt{slicelet} model. A \texttt{slicelet} encapsulates the type
of service to be consumed and the duration of consumption. Services are
constructed from a collection of Slices, each having a priority and a
weightage factor, representing a percentage of Network Function (NF)
resource allocation against that particular slice. Slicenet allows for
the creation of both static and dynamic slices, providing a mechanism
for operators to model user behavior consistently and compare various
optimization model's performances. Static slices are established at the
topology creation stage by the operator, whereas dynamic slices are
generated by the slice management system on-demand based on the traffic
pattern, load and estimations.

\hypertarget{managers}{%
\subsection*{MANAGERS}\label{managers}}
\addcontentsline{toc}{subsection}{MANAGERS}

\begin{figure}

{\centering \includegraphics{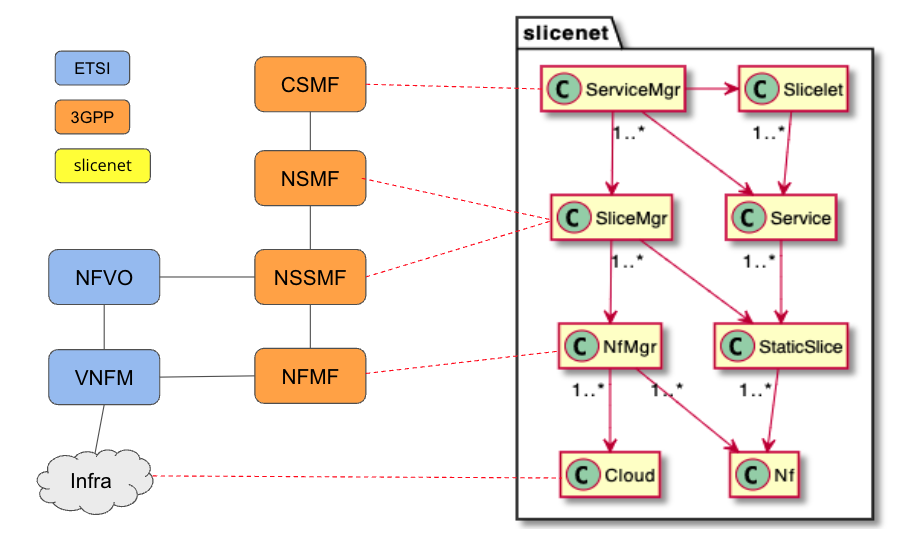}

}

\caption{\label{fig-slicenet-etsi-sa}slicenet package classes mapped to
ETSI MANO, 3GPP SA reference architecture}

\end{figure}

\texttt{NfMgr} class represents the \textbf{vNFM} functionality in
\emph{ETSI MANO architecture} and \textbf{NFMF} functionality in
\emph{3GPP SA Network Slice architecture}. This python static class
offers methods to register one or more Cloud objects, allocate,
schedule, deploy one or more Nf objects into the underlying Cloud
objects and to undeploy, unregister the entities respectively. This
class also offers hooks to influence the NF management operations by
specifying policies. A researcher or operator can create various slice
aware cloud and NF placement policies and study the behaviour to select
appropriate policy to their needs and usecases.

\texttt{SliceMgr} class represents the \textbf{NSMF/NSSMF} functionality
in \emph{3GPP SA Network Slice architecture}. In the first release of
the slicenet, this class abstracts the overall functionality of both
NSMF and NSSMF but future releases will offer seperate classes to better
represent the complexities and usecases between inter NSMF-NSSMF
coordination, dependency management and scheduling problem statements.
This python static class offers methods to compose one or more
StaticSlice object and to deploy and undeploy the slice instances.

\texttt{ServiceMgr} class represents the \textbf{CSMF} functionality in
\emph{3GPP SA Network Slice architecture}. This python static class
provides methods to compose one or more Service objects and to deploy
and undeploy the service instances. Figure~\ref{fig-slicenet-etsi-sa}
provides the mapping view of slicenet package classes to 3GPP SA / ETSI
MANO Reference architecture.

\hypertarget{sec-4}{%
\section{Simulating with Slicenet}\label{sec-4}}

\begin{figure}

{\centering \includegraphics{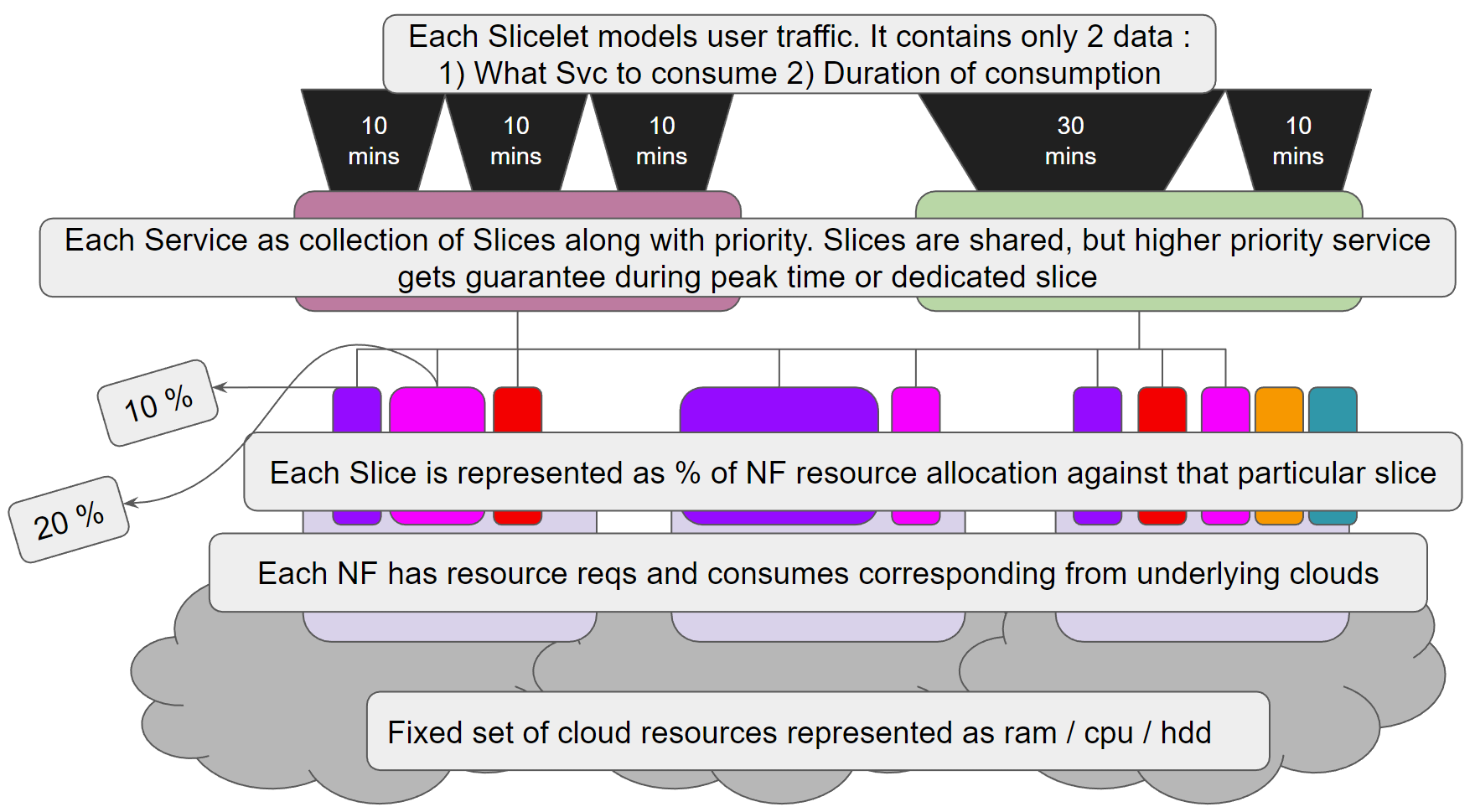}

}

\caption{\label{fig-slicenet-view}High level view of Slicelets consuming
the computing and networking resources of a underlying slice
infrastructure}

\end{figure}

To experiment with Slicenet, one must follow a series of steps from
creating Cloud objects to removing slices and tearing down the system.
Slicenet follows the typical create, set, fire method of using a python
class, similar to mininetlike experience. A pseudo code for creating
experiments with slicenet pypackage is provided in
Listing~\ref{lst-pseudo}. Figure~\ref{fig-slicenet-view} provides a
pictorial representation of how end user's traffic worloads modelled as
slicelets, gets applied to the underlying slicing infrastructure.

\begin{codelisting}

\caption{Pseudo Code for Slicenet Experiment}

\hypertarget{lst-pseudo}{%
\label{lst-pseudo}}%
\begin{Shaded}
\begin{Highlighting}[numbers=left,,]
\NormalTok{Create Cloud objects}
  \ControlFlowTok{for}\NormalTok{ each cloud }\BuiltInTok{object}\NormalTok{ do}
\NormalTok{    Register }\ControlFlowTok{with}\NormalTok{ NfMgr}
\NormalTok{  end }\ControlFlowTok{for}
\NormalTok{Set Nf Scheduler Policy}
\NormalTok{Create Nf objects}
\ControlFlowTok{for}\NormalTok{ each Nf }\BuiltInTok{object}\NormalTok{ do}
\NormalTok{  Deploy this }\BuiltInTok{object}\NormalTok{ using NfMgr Deploy method}
\NormalTok{end }\ControlFlowTok{for}
\NormalTok{Create StaticSlice objects}
\ControlFlowTok{for}\NormalTok{ each }\BuiltInTok{slice} \BuiltInTok{object}\NormalTok{ do}
\NormalTok{  deploy this }\BuiltInTok{slice}\NormalTok{ using SliceMgr deploy method}
\NormalTok{end }\ControlFlowTok{for}
\end{Highlighting}
\end{Shaded}

\end{codelisting}

A sample python listing of using the slicenet classes is provided in
Listing~\ref{lst-listing1}

\hypertarget{potential-experiments-powered-by-slicenet}{%
\subsection*{\texorpdfstring{\emph{Potential Experiments powered by
slicenet}}{Potential Experiments powered by slicenet}}\label{potential-experiments-powered-by-slicenet}}
\addcontentsline{toc}{subsection}{\emph{Potential Experiments powered by
slicenet}}

Potential experiments include exploring Slice composability, Slice aware
scheduling or placement, optimized Sub Slice Selection, Service to Slice
Selection, Dynamic Slicing, Dynamic QoS negotiation, Slice policy
generation and optimization, and Slice to resource mapping. Slicenet
also promotes the use of popular Python frameworks like TensorFlow or
Py-Torch for both training models and generating datasets for the
experiments. After the slicelet duration, the state of the network
objects can be inspected to report on utilization, SLA adherence,
composition, scheduling inference, and other metrics.

\hypertarget{benefits-of-using-slicenet-for-communication-service-providers-csps}{%
\subsection*{\texorpdfstring{\emph{Benefits of Using Slicenet for
Communication Service Providers
(CSPs)}}{Benefits of Using Slicenet for Communication Service Providers (CSPs)}}\label{benefits-of-using-slicenet-for-communication-service-providers-csps}}
\addcontentsline{toc}{subsection}{\emph{Benefits of Using Slicenet for
Communication Service Providers (CSPs)}}

Slicenet offers significant advantages to service providers. By
simulating their network, they can achieve valuable insights into
resource optimization, capacity planning, QoS assessment, cost
optimization, performance comparison, risk mitigation, and Service Level
Agreement (SLA) compliance. The gained insights and informed decisions
can help improve resource efficiency, enhance QoS, optimize costs, and
ensure reliable service delivery to their customers.

\hypertarget{sec-5}{%
\section{Results and Data Visualizations with Pythonic
Extensions}\label{sec-5}}

\begin{figure}

{\centering \includegraphics{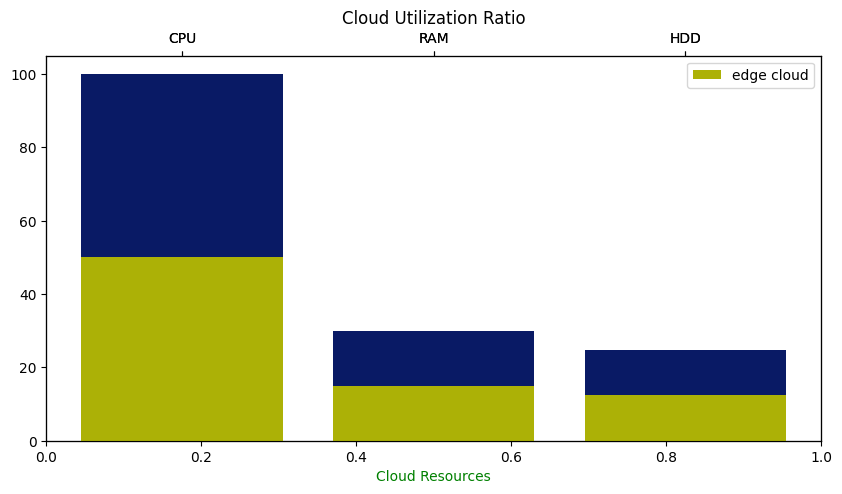}

}

\caption{\label{fig-cloud-util-ratio}Sample Cloud utilization ratio
reported by slicenet after deploying NFs}

\end{figure}

\begin{figure}

{\centering \includegraphics{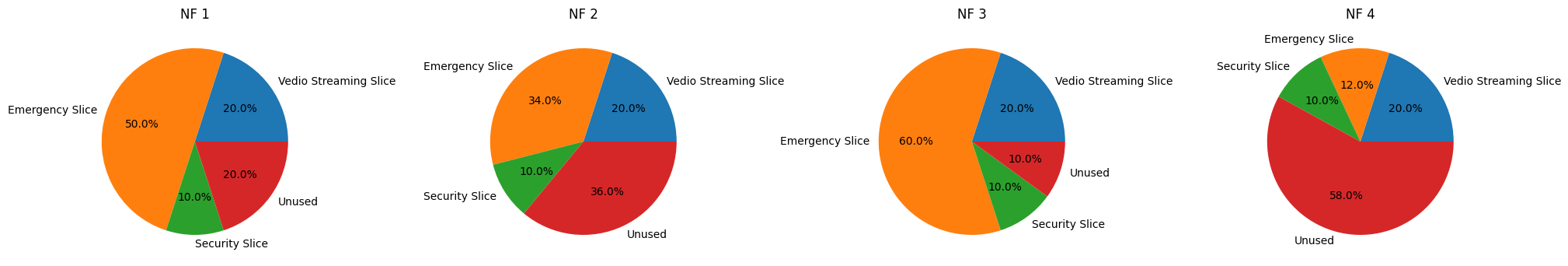}

}

\caption{\label{fig-slice-util-ratio}Sample Slice Utilization ratio on a
NF Level after adding slices to NFs}

\end{figure}

\begin{figure}

{\centering \includegraphics{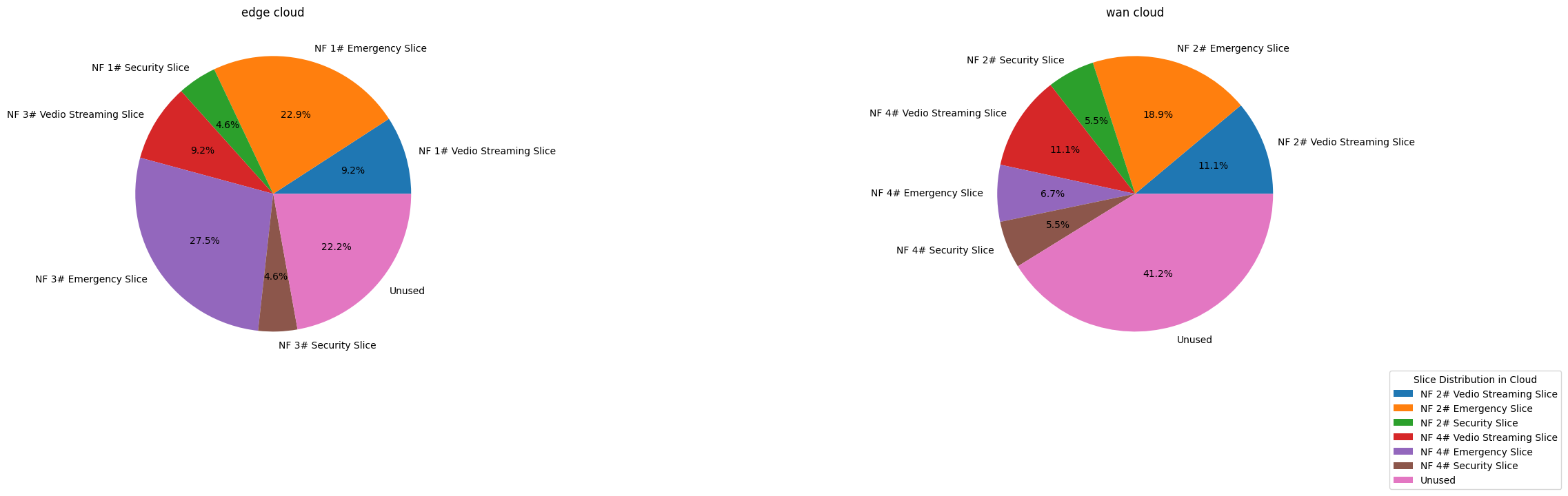}

}

\caption{\label{fig-slice-cloud-ratio}Sample 3 Layer view to understand
how a service is composed of slices, which inturn composed of NFs as
seen from Cloud Level}

\end{figure}

Apart from basic topology based experiments, slicenet can also be used
in more pythonic way. For example, slicenet can be used to record the
states of Nf, Cloud, StaticSlice etc objects and the before and after
state changes can be visualized using popular python packages like
matplotlib. Some of the useful visualizations for example are : Cloud
utilization ratio, Slice Composiblity view, Service to Slice to Nf to
Cloud level view etc. These visualizations provide powerful
understanding of the various optimization models for a researcher to
understand the effectiveness of the model. Also a communication service
provider can understand the effects of various models in their
underlying infrastructure thereby enabling them to select apt models for
their usecases. Listing~\ref{lst-listing2} provides a sample of how
various various visualization can be performed with popular python
package and Figure~\ref{fig-cloud-util-ratio},
Figure~\ref{fig-slice-cloud-ratio}, Figure~\ref{fig-slice-util-ratio}
are the output visualizations from Listing~\ref{lst-listing2}

\hypertarget{sec-6}{%
\section{Conclusion}\label{sec-6}}

Slicenet offers an efficient way for operators and researchers to
experiment with network slicing. Unlike existing tools, Slicenet allows
for a high degree of flexibility and customization in experiments,
making it a powerful tool for studying network slicing problem space.
Furthermore, by simulating networks rather than emulating them, Slicenet
is less resource intensive, enabling more extensive, complex and large
scale simulations.

This tool's benefits extend to service providers, who can gain valuable
insights into resource utilization, capacity planning, QoS, and cost
optimization. Slicenet is poised to deliver measurable, reproducible
results consistently, and to abstract traffic patterns from topology,
allowing identical topologies to be tested under varying traffic
conditions. In this light, Slicenet stands as a pivotal tool in the
realm of network slice management and orchestration, aiding in
overcoming the barriers and seizing the opportunities of network slicing
for 5G and beyond.

\newpage{}

\hypertarget{references}{%
\section*{References}\label{references}}
\addcontentsline{toc}{section}{References}

\hypertarget{refs}{}
\begin{CSLReferences}{0}{0}
\leavevmode\vadjust pre{\hypertarget{ref-DAN21}{}}%
\CSLLeftMargin{1. }%
\CSLRightInline{Chahbar M, Diaz G, Dandoush A, Cerin C, Ghoumid K (2021)
A comprehensive survey on the E2E 5G network slicing model. IEEE
Transactions on Network and Service Management 18(1):49--62.
\url{https://doi.org/10.1109/TNSM.2020.3044626}}

\leavevmode\vadjust pre{\hypertarget{ref-cnsmDAN}{}}%
\CSLLeftMargin{2. }%
\CSLRightInline{Chahbar M, Diaz G, Dandoush A (2019)
\href{https://doi.org/10.23919/CNSM46954.2019.9012745}{Towards a unified
network slicing model}. In: 2019 15th international conference on
network and service management (CNSM). pp 1--5}

\leavevmode\vadjust pre{\hypertarget{ref-ITASDAN}{}}%
\CSLLeftMargin{3. }%
\CSLRightInline{Kammoun A, Dandoush A, Diaz G, achir nadjib, Tabbane N
(2023) \href{https://doi.org/10.36227/techrxiv.22581145.v1}{Enabling new
generation services with multi-level delegation architecture of slicing
technology}. In: Accepted in ITAS 2023, doha, UDSt, 2023}

\leavevmode\vadjust pre{\hypertarget{ref-etsisdnfv}{}}%
\CSLLeftMargin{4. }%
\CSLRightInline{(2015) Network functions virtualisation (NFV);
ecosystem; report on SDN usage in NFV architectural framework. ETSI GS
NFV-EVE 005}

\leavevmode\vadjust pre{\hypertarget{ref-norma}{}}%
\CSLLeftMargin{5. }%
\CSLRightInline{Rost P, Mannweiler C, Michalopoulos DS, et al (2017)
Network slicing to enable scalability and flexibility in 5G mobile
networks. IEEE Communications Magazine 55(5):72--79.
\url{https://doi.org/10.1109/MCOM.2017.1600920}}

\leavevmode\vadjust pre{\hypertarget{ref-sonata}{}}%
\CSLLeftMargin{6. }%
\CSLRightInline{Dräxler S, Karl H, Peuster M, et al (2017)
\href{https://doi.org/10.1109/ICCW.2017.7962785}{SONATA: Service
programming and orchestration for virtualized software networks}. In:
2017 IEEE international conference on communications workshops (ICC
workshops). pp 973--978}

\leavevmode\vadjust pre{\hypertarget{ref-DAN23}{}}%
\CSLLeftMargin{7. }%
\CSLRightInline{Uddin M, Manickam S, Ullah H, Obaidat M, Dandoush A
(2023) Unveiling the metaverse: Exploring emerging trends, multifaceted
perspectives, and future challenges. IEEE Access 1--1.
\url{https://doi.org/10.1109/ACCESS.2023.3281303}}

\leavevmode\vadjust pre{\hypertarget{ref-DAN2010}{}}%
\CSLLeftMargin{8. }%
\CSLRightInline{Dandoush A, Jean-Marie A (2010)
\href{https://doi.org/10.1109/CTRQ.2010.23}{Flow-level modeling of
parallel download in distributed systems}. In: 2010 third international
conference on communication theory, reliability, and quality of service.
pp 92--97}

\leavevmode\vadjust pre{\hypertarget{ref-mininet}{}}%
\CSLLeftMargin{9. }%
\CSLRightInline{Team M (2012) \href{http://mininet.org}{Mininet an
instant virtual network on your laptop (or other PC)}}

\leavevmode\vadjust pre{\hypertarget{ref-Mininet-WiFi}{}}%
\CSLLeftMargin{10. }%
\CSLRightInline{Fontes RR, Afzal S, Brito SHB, Santos MAS, Rothenberg CE
(2015) \href{https://doi.org/10.1109/CNSM.2015.7367387}{Mininet-WiFi:
Emulating software-defined wireless networks}. In: 2015 11th
international conference on network and service management (CNSM). pp
384--389}

\leavevmode\vadjust pre{\hypertarget{ref-Containernet}{}}%
\CSLLeftMargin{11. }%
\CSLRightInline{Peuster M, Kampmeyer J, Karl H (2018)
\href{https://doi.org/10.1109/NETSOFT.2018.8459905}{Containernet 2.0: A
rapid prototyping platform for hybrid service function chains}. In: 2018
4th IEEE conference on network softwarization and workshops (NetSoft).
pp 335--337}

\leavevmode\vadjust pre{\hypertarget{ref-EdgeCloudSim}{}}%
\CSLLeftMargin{12. }%
\CSLRightInline{Sonmez C, Ozgovde A, Ersoy C (2017)
\href{https://doi.org/10.1109/FMEC.2017.7946405}{EdgeCloudSim: An
environment for performance evaluation of edge computing systems}. In:
2017 second international conference on fog and mobile edge computing
(FMEC). pp 39--44}

\leavevmode\vadjust pre{\hypertarget{ref-CloudSimHypervisor}{}}%
\CSLLeftMargin{13. }%
\CSLRightInline{Nyanteh AO, Li M, Abbod MF, Al-Raweshidy H (2021)
CloudSimHypervisor: Modeling and simulating network slicing in
software-defined cloud networks. IEEE Access 9:72484--72498.
\url{https://doi.org/10.1109/ACCESS.2021.3079501}}

\end{CSLReferences}

\newpage{}

\begin{codelisting}

\caption{Python example using slicenet}

\hypertarget{lst-listing1}{%
\label{lst-listing1}}%
\begin{Shaded}
\begin{Highlighting}[numbers=left,,]
\ImportTok{from}\NormalTok{ cloud }\ImportTok{import}\NormalTok{ Cloud}
\ImportTok{from}\NormalTok{ nf }\ImportTok{import}\NormalTok{ Nf}
\ImportTok{from}\NormalTok{ nfMgr }\ImportTok{import}\NormalTok{ NfMgr}
\ImportTok{from}\NormalTok{ sliceMgr }\ImportTok{import}\NormalTok{ SliceMgr}
\ImportTok{from}\NormalTok{ staticslice }\ImportTok{import}\NormalTok{ StaticSlice}
\ImportTok{from}\NormalTok{ tabulate }\ImportTok{import}\NormalTok{ tabulate}
\ImportTok{import}\NormalTok{ random}

\NormalTok{clouds }\OperatorTok{=}\NormalTok{ []}
\NormalTok{nfs }\OperatorTok{=}\NormalTok{ []}
\NormalTok{clouds.append(Cloud(}\DecValTok{1000}\NormalTok{, }\DecValTok{10}\NormalTok{, }\DecValTok{10000}\NormalTok{, name}\OperatorTok{=}\StringTok{"c1"}\NormalTok{))}
\NormalTok{clouds.append(Cloud(}\DecValTok{2000}\NormalTok{, }\DecValTok{20}\NormalTok{, }\DecValTok{20000}\NormalTok{, name}\OperatorTok{=}\StringTok{"c2"}\NormalTok{))}

\CommentTok{\#\# Static Slices}
\NormalTok{nf1 }\OperatorTok{=}\NormalTok{ Nf(’NF }\DecValTok{1}\NormalTok{’, }\DecValTok{100}\NormalTok{,}\DecValTok{9}\NormalTok{,}\DecValTok{1234}\NormalTok{)}
\NormalTok{nf2 }\OperatorTok{=}\NormalTok{ Nf(’NF }\DecValTok{2}\NormalTok{’, }\DecValTok{100}\NormalTok{,}\DecValTok{9}\NormalTok{,}\DecValTok{1234}\NormalTok{)}
\NormalTok{nf3 }\OperatorTok{=}\NormalTok{ Nf(’NF }\DecValTok{3}\NormalTok{’, }\DecValTok{200}\NormalTok{,}\DecValTok{1}\NormalTok{,}\DecValTok{1234}\NormalTok{)}
\NormalTok{nf4 }\OperatorTok{=}\NormalTok{ Nf(’NF }\DecValTok{4}\NormalTok{’, }\DecValTok{200}\NormalTok{,}\DecValTok{1}\NormalTok{,}\DecValTok{1234}\NormalTok{)}

\NormalTok{nfs.append(nf1)}
\NormalTok{nfs.append(nf2)}
\NormalTok{nfs.append(nf3)}
\NormalTok{nfs.append(nf4)}

\NormalTok{NfMgr.setSchedularPolicy(’first}\OperatorTok{{-}}\NormalTok{available}\OperatorTok{{-}}\NormalTok{method’)}

\ControlFlowTok{for}\NormalTok{ c }\KeywordTok{in}\NormalTok{ clouds:}
\NormalTok{  NfMgr.registerCloud(c)}
  \BuiltInTok{print}\NormalTok{(}\SpecialStringTok{f"Deploying NFs as a planned activity : Static}
\ErrorTok{Slices")}

\ControlFlowTok{for}\NormalTok{ n }\KeywordTok{in}\NormalTok{ nfs:}
\NormalTok{  NfMgr.deployNf(n)}

\CommentTok{\#print(f"After Deployment")}
\CommentTok{\#print("Cloud Utilization Ratio")}
\NormalTok{NfMgr.dumpCloudInfo()}
\CommentTok{\#print("NF to Cloud Allocation Ratio")}
\BuiltInTok{print}\NormalTok{(}\SpecialStringTok{f"Before Adding Slices"}\NormalTok{)}
\NormalTok{NfMgr.dumpNfInfo()}

\CommentTok{\# Adding Static Slices}
\NormalTok{vs\_slice }\OperatorTok{=}\NormalTok{ StaticSlice(}\StringTok{"Vedio Streaming Slice"}\NormalTok{)}
\NormalTok{vs\_slice.composeSlice(nf1.}\BuiltInTok{id}\NormalTok{, }\DecValTok{20}\NormalTok{)}
\NormalTok{vs\_slice.composeSlice(nf2.}\BuiltInTok{id}\NormalTok{, }\DecValTok{20}\NormalTok{)}
\NormalTok{vs\_slice.composeSlice(nf3.}\BuiltInTok{id}\NormalTok{, }\DecValTok{20}\NormalTok{)}
\NormalTok{vs\_slice.composeSlice(nf4.}\BuiltInTok{id}\NormalTok{, }\DecValTok{20}\NormalTok{)}
\NormalTok{es\_slice }\OperatorTok{=}\NormalTok{ StaticSlice(}\StringTok{"Emergency Slice"}\NormalTok{)}
\NormalTok{es\_slice.composeSlice(nf1.}\BuiltInTok{id}\NormalTok{, }\DecValTok{50}\NormalTok{)}
\NormalTok{es\_slice.composeSlice(nf2.}\BuiltInTok{id}\NormalTok{, }\DecValTok{34}\NormalTok{)}
\NormalTok{es\_slice.composeSlice(nf3.}\BuiltInTok{id}\NormalTok{, }\DecValTok{60}\NormalTok{)}
\NormalTok{es\_slice.composeSlice(nf4.}\BuiltInTok{id}\NormalTok{, }\DecValTok{12}\NormalTok{)}
\NormalTok{SliceMgr.deploySlice(vs\_slice)}
\NormalTok{SliceMgr.deploySlice(es\_slice)}
\NormalTok{SliceMgr.dumpSlices()}

\BuiltInTok{print}\NormalTok{(}\SpecialStringTok{f"After Adding Random Slices"}\NormalTok{)}
\BuiltInTok{print}\NormalTok{(}\SpecialStringTok{f"After adding static slices"}\NormalTok{)}
\NormalTok{NfMgr.dumpNfInfoSliceDetails()}
\end{Highlighting}
\end{Shaded}

\end{codelisting}

\newpage{}

\begin{codelisting}

\caption{IPython / Jupyter notebook example using slicenet}

\hypertarget{lst-listing2}{%
\label{lst-listing2}}%
\begin{Shaded}
\begin{Highlighting}[numbers=left,,]
\ImportTok{import}\NormalTok{ matplotlib.pyplot }\ImportTok{as}\NormalTok{ plt}
\NormalTok{labels }\OperatorTok{=}\NormalTok{ [}\StringTok{"Video Streaming Slice"}\NormalTok{, }\StringTok{"Emergency Slice"}\NormalTok{, }\StringTok{"Unused"}\NormalTok{]}
\NormalTok{fig }\OperatorTok{=}\NormalTok{ plt.figure(figsize}\OperatorTok{=}\NormalTok{(}\DecValTok{10}\NormalTok{, }\DecValTok{6}\NormalTok{))}
\NormalTok{sep }\OperatorTok{=} \DecValTok{0}
\ControlFlowTok{for}\NormalTok{ n }\KeywordTok{in}\NormalTok{ nfs:}
\NormalTok{  weights }\OperatorTok{=}\NormalTok{ []}
\NormalTok{  labels }\OperatorTok{=}\NormalTok{ []}
  \ControlFlowTok{for}\NormalTok{ k,v }\KeywordTok{in}\NormalTok{ n.slices.items():}
\NormalTok{    weights.append(v)}
\NormalTok{    labels.append(SliceMgr.slices[k].name)}
\NormalTok{  weights.append(n.getSliceCount())}
\NormalTok{  labels.append(}\StringTok{"Unused"}\NormalTok{)}
\NormalTok{  ax\_ }\OperatorTok{=}\NormalTok{ fig.add\_axes([}\DecValTok{0}\OperatorTok{+}\NormalTok{sep, }\DecValTok{0}\NormalTok{, }\FloatTok{.5}\NormalTok{, }\FloatTok{.5}\NormalTok{], aspect}\OperatorTok{=}\DecValTok{1}\NormalTok{)}
\NormalTok{  ax\_.pie(weights, labels}\OperatorTok{=}\NormalTok{labels, radius }\OperatorTok{=} \DecValTok{1}\NormalTok{, autopct}\OperatorTok{=}\NormalTok{’}\OperatorTok{\%}\FloatTok{1.1}\ErrorTok{f}\OperatorTok{\%\%}\NormalTok{’)}
\NormalTok{  ax\_.set\_title(n.name)}
\NormalTok{  sep }\OperatorTok{+=} \FloatTok{0.5}
\NormalTok{plt.show()}
\end{Highlighting}
\end{Shaded}

\end{codelisting}

\end{document}